\documentclass[twocolumn]{article}
\usepackage{epsfig}

\newcommand{\inputEps}[3]{%
         \vskip 12pt
            \centerline{\epsfxsize=#1 \epsfbox{#3} }
            \vskip 12pt
            {\center\small{#2}}
            \vskip 18pt
            }

\begin{document}

\title{Free-Electron Lasers Without Inversion: 
       Design of Two-Magnet Drift Region}

\date{}
\author{
\begin{minipage}{0.7\textwidth}
\centerline{A.I. Artemyev$^{(1,2,3)}$, Yu.V. Rostovtsev$^{(2)}$,}
\vskip 1mm
\centerline{S. Trendafilov$^{(2)}$, K. Kapale$^{(2)}$, }
\vskip 1mm
\centerline{M.V. Fedorov$^{(3)}$, G. Kurizki$^{(1)}$,}
\vskip 1mm
\centerline{M.O. Scully$^{(2)}$}
\vskip 3mm
\centerline{
$^{(1)}$Department of Chemical Physics,
Weizmann Institute of Science,
76100 Rehovot, Israel}
\vskip 1mm
\centerline{$^{(2)}$Department of Physics,
Texas A\&M University, College Station,
Texas, 77843-4242, USA}
\vskip 1mm
\centerline{$^{(3)}$General Physics Institute,
38 Vavilov St.,
Moscow, 119991, Russia
}
\end{minipage}
}

\maketitle

\begin{abstract}
We propose a two-magnet design of a drift region for a
free-electron laser without inversion (FELWI). By performing
direct calculations of the phase shifts for electrons passing the
drift region, we prove that the small-signal gain integrated over
the detuning is positive and is inversely proportional to the
energy spread of the ``hot'' electron beam. The dispersion and the
geometry of the drift region are specified, and the requirements
to the electron beam quality, including the transverse size and
the angular spread, are found.
\end{abstract}



\section{Introduction}

Free electron lasers (FELs) are able to produce radiation in
different domains, from microwaves \cite{long-wavelength-FELs} to
X-rays \cite{short-wavelength-FELs}. They have found many
applications \cite{FELs-scientific-applications}, including
nonlinear spectroscopic characterization of quantum wells
\cite{FEL-quantum-wells-spectrosopy}, solid surfaces
\cite{nonlinear-spectroscopy-of-solid-films}, for near-field
surface microscopy \cite{near-field-surface-microscopy}, infrared
photodissociation spectroscopy of molecules
\cite{Molecular-spectroscopy-FELs}, laser surgery
\cite{Laser-surgery}, \cite{Laser-surgery-eye}, material research
and processing \cite{FEL-laser-technology}, interaction of X-rays
with matter \cite{FEL-X-ray-matter-interaction}, etc. The FEL gain
is attributed to the interference of the amplified electromagnetic
wave $E_0$ and the waves emitted by the electrons at the wiggler
magnets. The phases of the wiggler-induced waves are sensitive to
the electron velocity detuning from the resonance. This leads to
the odd gain profile shown in Fig.~1. The the constructive
interference, which is needed for the positive gain, requires
matching these phases and impose limitations on the electron beam
energy spread, angular divergence, transverse size, and,
consequently on the maximum FEL frequency and power
\cite{FEL-emittance}.

\vskip 12pt

\inputEps{220pt}{Figure 1. Interference of waves emitted at wiggler
magnets and gain of an FEL.}{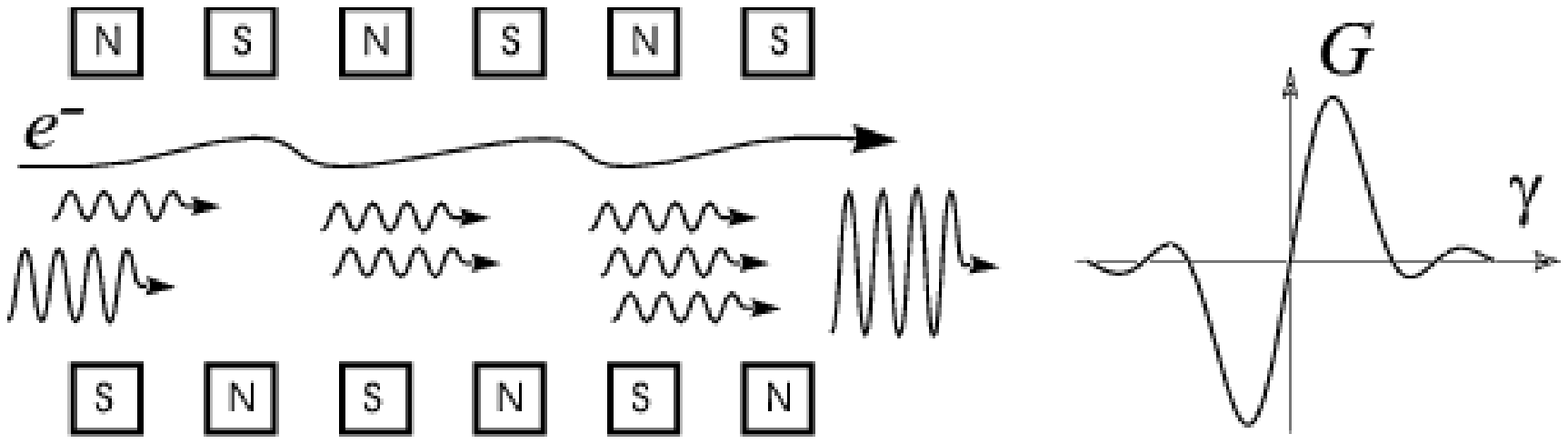}

A laser without inversion (LWI) \cite{LWI} relies on the phase
control of the electron transition amplitudes, interference
suppression of the stimulated absorption and keeping intact the
stimulated emission. This makes the gain possible even if there is
no inversion, i.e., if the majority of electrons are in the lower
energy states. The LWI phase control is achieved by the combined
action of the external driving laser and the spontaneous
relaxation. A new kind of FEL was proposed \cite{1}, which
implements ideas of lasing without inversion. In an free-electron
laser without inversion (FELWI) interfering free-free electron
transition amplitudes are formed in wigglers, which are spatially
separated by the drift region. The drift region delays different
electrons at different phase shifts $\Delta\psi_D$ relative to the
amplified wave. For each electron a proper phase delay can lead to
a constructive interference of the amplified wave $E_0$ and the
radiation emitted in the two wigglers. Thus the gain profile of an
FELWI is positive for all energies, making lasing possible even if
there is no inversion, i.e., if there is an equal amount of
electrons above and below the resonant energy. The positive
small-signal gain profile shown in Fig.~2 makes an FELWI
advantageous over an optical klystron or a one-wiggler FEL: for a
broad width $\Delta {\cal E}$ of the electron energy distribution,
an FELWI gain scales as $\sim \; 1/\Delta {\cal E}$ and exceeds
the gain of a one-wiggler FEL or an optical klystron gain, which
scale as $\sim \; 1/\Delta {\cal E}^2$. Thus an FELWI can be
considered as an optical klystron with an improved phase control
or a ``phased'' optical klystron.

  \centerline{
  \vphantom{ \epsfig{file=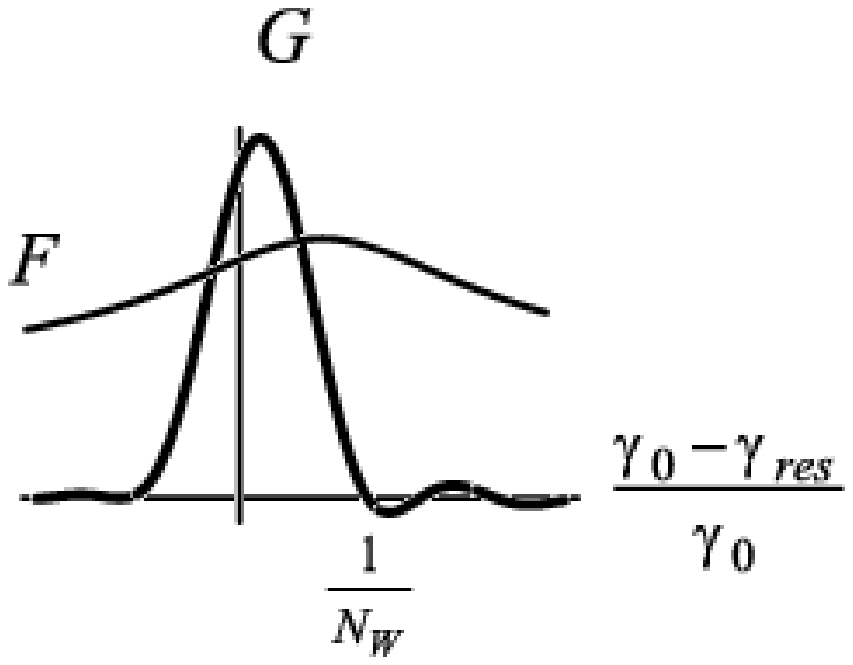,  width=3.7cm, height=3.6cm, angle=0}}
  \epsfig{file=./fig3gainfelwi-cdr10.eps,  width=3.7cm, angle=0}
  \epsfig{file=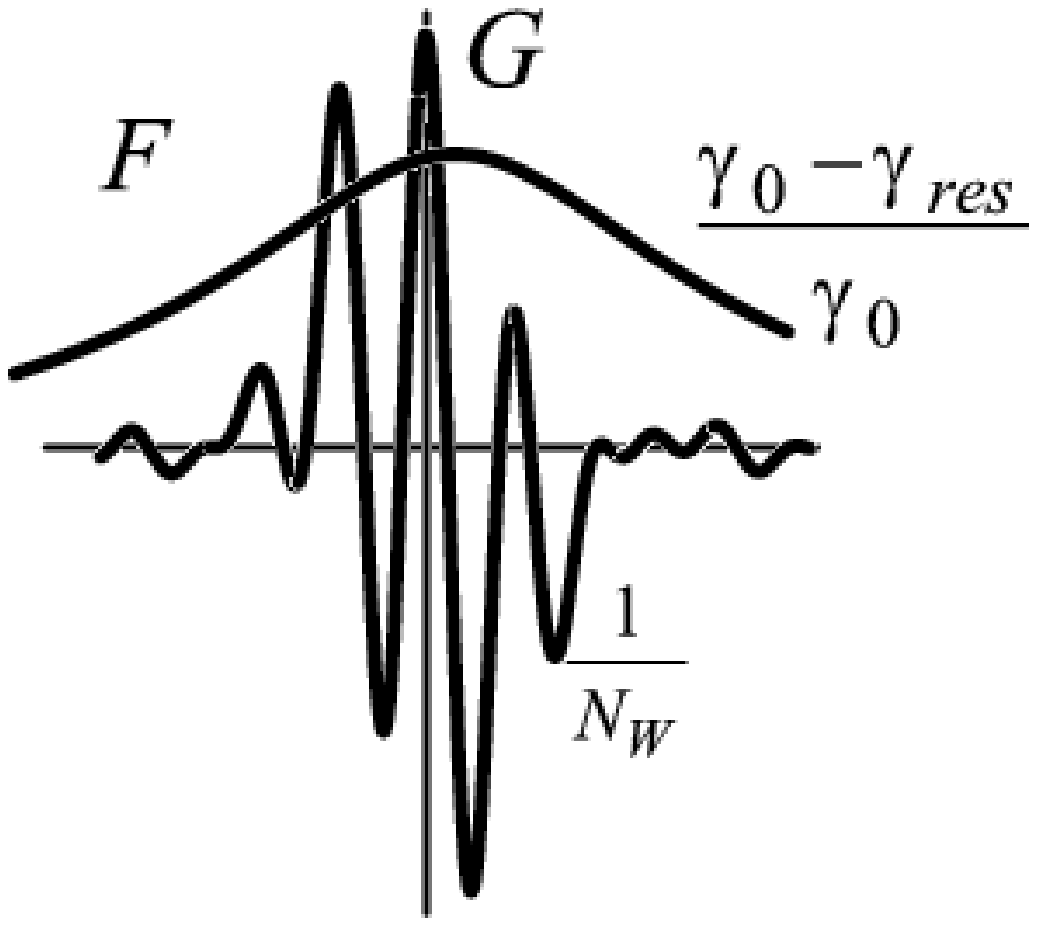,  width=3.3cm, height=2.8cm, angle=0}
  }
  {\small{Figure 2. FELWI versus optical klystron gain profile.}}

\vskip 18pt

The physics of an FELWI was studied in many important details
\cite{1}--\cite{6}. It was shown that a FELWI requires a
correlation between the transverse electron velocity and the
electron energy change in the first wiggler \cite{4}. A
non-collinear FEL geometry provides this correlation and allows
for the essentially two-dimensional electron dynamics in the drift
region, which is needed for the FELWI phase control \cite{4},
\cite{5}. It was shown that the FELWI is consistent with the
Liouville's and the generalized Madey's theorems \cite{5}. In this
sense a multi-dimensional electron dynamics in an FELWI drift
region is the classical analog of the spontaneous relaxation in a
quantum LWI. Different kinds of drift region phase shifts were
studied, including step-like phase shifts \cite{2}--\cite{5} and a
linear phase shift \cite{6}. An electron optics design for the
linear phase shift was proposed \cite{6}.

In this paper, we suggest a two-magnet drift region for an FEL
without inversion and prove by direct calculation of electron
motion in the drift region that the suggested design indeed gives
us FELWI gain. Based on the suggested drift region, we are able to
address many important questions of FELWI physics, which remained
unsolved: how sensitive is the FELWI phase control to the angular
spread of the electron beam, and how much the gain can be enhanced
with the phase shifts implemented with the conventional electron
optics elements.

The paper is outlined as follows. In Section 2 we describe the
electron and field dynamics in the wigglers and find correlations
between the electron energy change in the wiggler and their
field-induced deflection. In Section 3 we describe the suggested
geometry of the drift region and by tracing electron trajectories
through the drift region we determine requirements to the electron
beam quality and estimate the minimum wavelength of the FEL. In
Section 4 we calculate the ``hot''-beam gain of the FELWI for an
arbitrary linear phase delay in the drift region. We find the
acceptable range of the FELWI drift region dispersion and
demonstrate how the gain changes with the increasing sensitivity
of the drift region phase shift to the field-induced changes of
the electron trajectory. Finally, in Section 5 we summarize our
results.

\section{Electron and field dynamics in a FELWI}

The classical electron dynamics in an FEL is described by
Hamiltonian

\begin{equation}
\label{hamiltonian-MKS} H \equiv \gamma m c^2 = \sqrt{
(\vec{p}-e\vec{A} )^2 c^2 + m^2 c^4},
\end{equation}
where $c$ is the speed of light, $e$, $m$, $\gamma$, and $\vec p$
are the electron charge, mass, Lorentz factor, and canonical
momentum, and

\begin{equation}
\label{vector-potential} \vec{A} \equiv \vec{e}_y A_y = \vec{e}_y
( A_L   e^{i\psi_L} + A_W   e^{i\psi_W} + \; C.C. )
\end{equation}
is the vector potential of the combined electromagnetic field of
the laser (designated by a subscript $L$) and the wiggler
(designated by a subscript $W$). The electron and laser beams
propagate at the small angles $\alpha$ and $\theta$ to the axis of
the wiggler, as in Fig.~3. The $z$-axis is directed along the
wiggler, so that $\psi_W=k_W z$, and the phase of the laser field
$\psi_L$ equals $k_L ( z \cos\theta + x \sin \theta)-\nu_L t +
\phi_0$.

\vskip 6pt

\inputEps{180pt}{Figure 3. Oblique geometry of an FEL.}{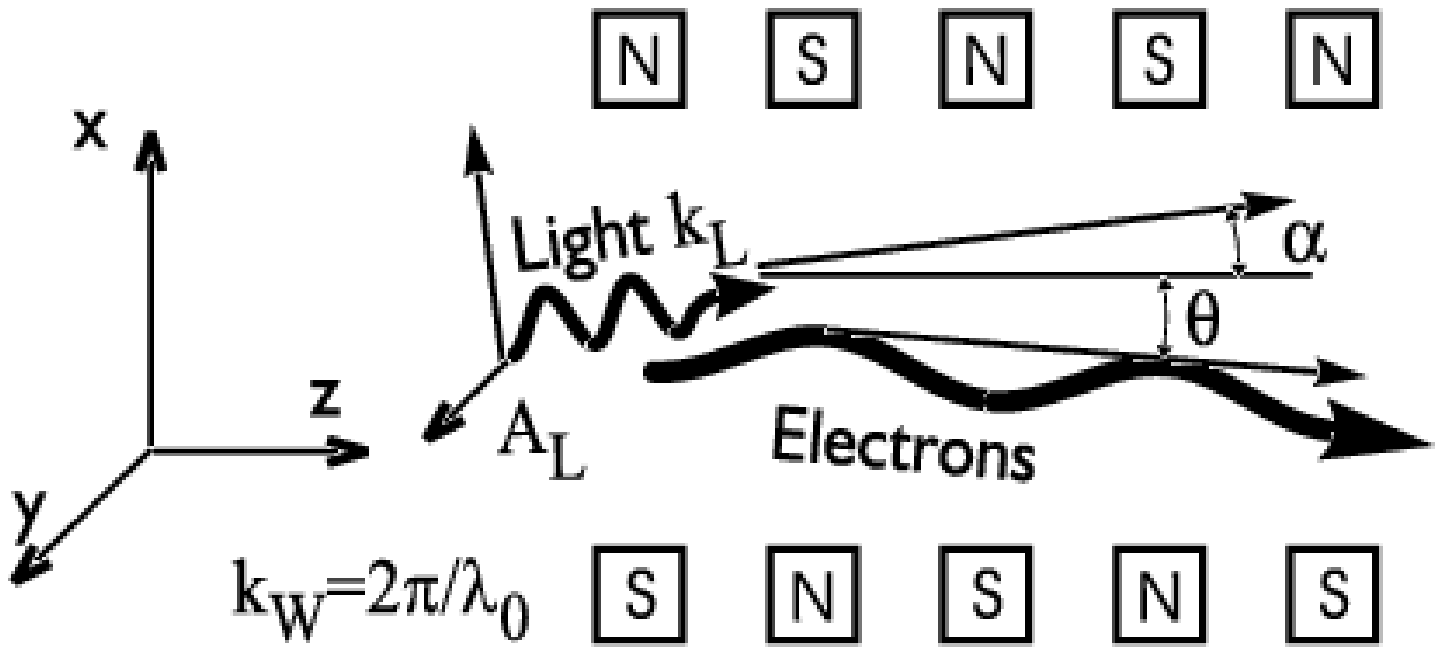}


The evolution of the slow phase $\psi$

\begin{equation}
 \label{psi-definition}
 \psi
   =
  \psi_L+\psi_W
   =
  (\vec k_L+\vec k_W)\cdot \vec r - \nu_L t + \phi_0
\end{equation}
is determined by the pendulum equation, which follows from
Eqs.~(\ref{dpxdtA2-appendix})--(\ref{dzdt1-appendix}) of the
Appendix:

\begin{eqnarray}
\label{pendulum-equation-system-dpsidt} \dot{\psi}
 & = & \Omega {\ }
 = {\ } (\vec k_L + \vec k_W)\cdot(\vec\beta-\vec\beta_{res})\,c,
\\
\label{pendulum-equation-system-dOmegadt} \ddot\psi &\equiv&
\dot{\Omega} {\ }={\ } \mbox{\ae}_0 \frac{A_L e^{i\psi} - A_L^*
e^{-i\psi}}{i},
\end{eqnarray}
where
\begin{equation}
\label{ae0} \mbox{\ae}_0 = \displaystyle \frac{2 e^2 A_W k_W
\nu_L}{m^2\gamma^2 c},
\end{equation}
and $\vec\beta_{res}\!=\vec{v}_{res}/c\!=\vec{e}_{\alpha}
\nu_L/(c\; \vec k_L \!\! \cdot\!\vec{\vphantom{k} e}_{\alpha} )$
is the dimensionless resonant electron velocity, $\vec e_{\alpha}
= (\sin\alpha, 0, \cos\alpha)$ is the unit vector along the
electron beam direction, $\vec \beta\equiv\vec
v/c=\{\beta_{x},0,\beta_{z}\}$ is the dimensionless electron
velocity in the wiggler determined by
Eqs.~(\ref{bety-wiggler-field-appendix})-(\ref{betz-wiggler-field-appendix}).
Initial conditions for
Eqs.~(\ref{pendulum-equation-system-dpsidt}),~(\ref{pendulum-equation-system-dOmegadt})
specify the slow phase $\psi_I(0)$ and $\psi_{II}(T)$ (for the
first and second wigglers respectively) and the resonant detuning
$\dot\psi_I(0)\equiv\Omega_0$ and
$\dot\psi_{II}(T)\equiv\Omega_1$:

\begin{eqnarray}
\label{pendulum-equation-system-initial-psi}
 \psi_I(0)&=&\phi_0,
\\ \label{Omega0-definition}
 \Omega_0 {\ }\equiv{\ }
 \dot\psi_I(0)&=& (\vec k_L+\vec k_W)\cdot \vec \beta_0 c - \nu_L,
\\
\label{pendulum-equation-system-initial-psi-2nd}
 \psi_{II}(T)&=&\psi_I(T)+\Delta\psi_D,
\\ \label{Omega0-definition-2nd}
 \Omega_1 {\;}\equiv{\;}
 \dot\psi_{II}(T)&=& \dot\psi_{I}(T).
\end{eqnarray}

We assume in
Eqs.~(\ref{pendulum-equation-system-initial-psi-2nd}),~(\ref{Omega0-definition-2nd})
that the role of the drift region is to introduce the phase shift
$\Delta\psi_D$ and that the resonant detuning $\Omega$ is not
changed in the drift region: $\dot\psi_{II}(T)=\dot\psi_{I}(T)$.
The change of the resonant detuning in the drift region can be
neglected if we assume that both the electron and laser beams
propagate in the second wiggler along its axis. In this case the
angular spread $\Delta\alpha_D$ cause negligible, second-order
corrections to the initial conditions.

For a non-collinear FEL geometry, $|\alpha|+|\theta|>0$, the
electron dynamics in the first wiggler reveals important
correlations. The changes of the resonant detuning
$\Delta\Omega_1=\Omega_I(T)-\Omega_I(0)\equiv\Omega_1-\Omega_0$
and the slow phase $\Delta\psi_1=\psi_I(T)-\psi_I(0)$ in the first
wiggler correlate with the wiggler-induced changes of the electron
energy $\Delta\gamma_1$, the transverse electron velocity $\Delta
v_{x\,1}$, the turn angle $\Delta\alpha_1$ of the electron
velocity in the first wiggler, and the transverse electron
position $\Delta x_1$:

\begin{eqnarray}
 \nonumber
 \Delta\gamma_1 &=& \frac{\gamma}{2 c k_W}
 \,\Delta\Omega_1,
\\ \label{Delta-gamma-alpha-x-Delta-Omega-1}
 \Delta\alpha_1 = \frac{\Delta v_{x\,1}}{c} &=&
 \frac{\theta-\alpha}{2 c k_W} \, \Delta\Omega_1,
\\
 \nonumber
 {\Delta x_{1}} &=& \frac{\theta-\alpha}{2 k_W} \, \Delta\psi_1.
\end{eqnarray}
We interpret these correlations in terms of the photon emission
and absorption: the electrons that absorb energy from the field
get deflected in the direction of the laser beam. Depending on
their initial conditions, the electrons can earn or loose energy.
The angular separation of the ``radiating'' and ``absorbing''
electrons makes it possible to treat these electrons differently
in the drift region. This provide an additional degree of freedom
in the phase control and allows one to arrange the FELWI phase
shift.


The pendulum equation
(\ref{pendulum-equation-system-dpsidt}),~(\ref{pendulum-equation-system-dOmegadt})
is solved by considering the electron-light interaction as a
perturbation and expanding the slow phase in the power series over
the laser vector potential,
$\psi_I(t)=\psi^{(0)}_{I}(t)+\psi^{(1)}_{I}(t)$ in the first
wiggler, $0<t<T$, and
$\psi_{II}(t)=\psi^{(0)}_{II}(t)+\psi^{(1)}_{II}(t)+ \Delta\psi_D$
in the second wiggler, $T<t<2T$. The zero-order approximation to
the phase $\psi$ is linear in time:
$\psi^{(0)}_I=\phi_0+\Omega_{0}t$ and $\psi^{(0)}_{II}=\phi_0+
\Delta\psi_D +\Omega_0 t$ in the first and second wigglers
respectively. The first-order corrections $\psi^{(1)}_I$ and
$\psi^{(1)}_{II}$ are found by treating the laser field as a
perturbation. The iterative procedure is described in
\cite{4},~\cite{5}. We assume that the laser frequency is close to
the resonant frequency $\nu_{res}$ determined by the condition
$\dot\psi=0$:

\begin{equation}
 \nu_{res}=
  \frac{2 c k_W\gamma^2}{1+K^2/2+\gamma^2(\alpha-\theta)^2}.
\end{equation}
The phase evolution is coupled to the change of the laser field
through the Maxwell's equations. The current density $\vec J=n_e
e\vec v$ and the electron velocity $\vec v$ are given by
Eq.~(\ref{hamilt-eqs-dydt-appendix}) of the Appendix. The wave
equation for the laser field envelope $A_L$ is simplified by the
slow varying envelope approximation, which allows to neglect
second derivatives of the field $A_L$, and the resonant
approximation, which ignores the fast-oscillating terms:

\begin{eqnarray}
\label{AL-Slow-Envelope}
 &&
    \left(
        \vec k_L \cdot \nabla  +
        \frac{\nu_L}{c^2}
        \,
        \frac{\partial }{\partial t}
    \right) A_L
    \,=\,
\\
\nonumber &&
 \,=\,
 \left< \frac{1}{2i} \, \frac{  \mu_0 n_e e^2
    (A_L +A_W^* e^{-i(\psi_L+\psi_W)}
    )}{m \gamma}
\right>
\\
\nonumber && \,\approx \, \left< \frac{N}{i}\, e^{-i \psi}
\right>,
\end{eqnarray}
where angular brackets denote averaging over electron distribution
function, $N= {\omega_P^2 A_W^*}/{\nu_L c \gamma} $, and
$\omega_P^2 = {\mu_0 n_e e^2 c^2 }/{ m}$, and
$\mu_0=4\pi\cdot10^{-7}\,\mathrm{N\cdot A^{-2}}$ is permeability
of vacuum (SI units). For a stationary FEL operation, $\partial
A_L/\partial t$=0, the change of the laser vector potential
$\Delta A_L$ is found by integrating Eq.~(\ref{AL-Slow-Envelope})
over the length of the two wigglers:

\begin{eqnarray}
\label{Delta-AL-two-integrals} && \Delta A_L \,=\, \left<
   \int\limits_0^{L_{W}}
   \frac{N}{i}
   \,
   e^{ -i(\psi^{(0)}+\psi^{(1)}) }
   \,dz
   \;+
\right.
\\ \nonumber
 &&
\left.
 \!\!
 +
\int\limits_{L_{W}}^{2L_{W}} \!\!
 \frac{N}{i}\,
 e^{-i(\psi^{(0)}+\psi^{(1)}+\Delta\psi_D)}
\,dz
 \right>.
\end{eqnarray}

For a two-wiggler FEL separated by a drift region, the change of
the vector potential $\Delta A_L$ of
Eq.~(\ref{Delta-AL-two-integrals}) depends on the initial phase
$\phi_0$ and the phase shift $\Delta\psi_D$ introduced by drift
region. A proper drift region phase shift $\Delta\psi_D$, which
can depend on the initial phase $\phi_0$, arranges the
constructive interference of the amplified radiation and the
radiation emitted in the two wigglers by most of electrons
regardless of their initial energy. In this papaer, we consider
implementation of an FELWI with  a linear phase shift

\begin{eqnarray}
\label{Delta-psi-D} \Delta \psi_D  =
    \varsigma
  + \chi \, \Omega_1 \, T \,
  + \xi  \, \Delta\Omega_1 \, T\,
  + \vartheta  \, \Delta\psi_1,
\end{eqnarray}
where $T=L_W/c$ is the time it takes electrons to pass the
wiggler. The coefficients $\varsigma$, $\chi$, $\xi$, and
$\vartheta$ are determined by the drift region geometry. They
describe the constant part of the phase shift, the drift region
dispersion $T^{-1}\partial \Delta\psi_D / \partial \Omega_1$, and
the sensitivity of the phase shift to the change of the resonant
detuning $\Delta\Omega_1$ and to the phase change $\Delta\psi_1$
in the first wiggler.

\section{FELWI drift region with two magnets}

The FELWI phase control can be achieved by making the path length
through the drift region sensitive to the electron energy and the
transverse electron motion in the wiggler. First,
 the electron path length should increase at large $\gamma_1$
 providing the negative dispersion
 of the drift region,  $\partial \Delta\psi_D/\partial\gamma_1<0$.
 This condition is necessary for any FELWI, and we specify
 the acceptable range of the drift region dispersion below.
It was shown \cite{4}--\cite{6} that the this requirement alone is
not enough for an FELWI and that the sensitivity of the drift
region to the transverse electron motion is essential for the free
electron lasing without inversion. In  this paper, we explore the
 phase shift $\Delta\psi_D$ which depends linearly
 on the direction of the electron velocity after the first wiggler,
 which is linear in $\Delta\Omega_1 $, and
 and on  the wiggler-induced shift of the electron transverse position $x_1$, which
 is proportional to the field-induced change
 of the resonant phase $\Delta\psi_1$ of
 Eq.~(\ref{Delta-gamma-alpha-x-Delta-Omega-1}).

For an FELWI to occur the linearized phase shift, which also
depends on the initial width $\Delta x_0$ and angular spread
$\Delta\alpha_0$ of the electron beam

\begin{eqnarray}
 \nonumber
 \Delta\psi
 & =  &
 \frac{\partial \Delta\psi_D}{\partial \gamma_1}\,(\gamma_1-\gamma_{res}) \,+\,
 \\  &&
 \frac{\partial \Delta\psi_D}{\partial \alpha}\,(\Delta \alpha_0+\Delta\alpha_1) \,+\,
 \\ \nonumber &&
 \frac{\partial \Delta\psi_D}{\partial x}\,(\Delta x_0+\Delta x_1) \,+\,
 const,
\end{eqnarray}
is to be matched to the linear phase shift of
Eq.~(\ref{Delta-psi-D}). Note that the initial width $\Delta x_0$
and initial angular spread $\Delta\alpha_0$ can wash out the FELWI
gain, as they contribute to the ``constant'' part of the phase
shift $\varsigma$. Thus the FELWI drift region can impose its own
limitations on the quality of the electron beam.

The suggested implementations of the FELWI phase shift
\cite{4}--\cite{7} assume manipulation of the electron beam with
the magnetic optics elements or the optical beam manipulation with
the set of prisms or a Bragg reflector. In this paper, for the
first time, we suggest and analyze in detail the performance of a
magnetic optics set for the FELWI drift region. The set consists
of two magnetic lenses which turn the electron beam in the same
direction. The sensitivity of angular and coordinate control is
achieved by aligning the first magnet almost parallel to the
electron beam at the exit from the first wiggler. Then the
electrons exiting the first wiggler at slightly different angles
meet the first magnet at rather different points and pass the
drift region along different routs. The negative dispersion is
achieved by letting the high-energy electron, which are less
deflected by the first magnet, pass the drift region along a
longer route, while lower-energy electrons, which are deflected by
the first magnet stronger, pass the drift region along the shorter
rout, as shown in Fig.~4.


{\epsfxsize=70mm \epsfbox{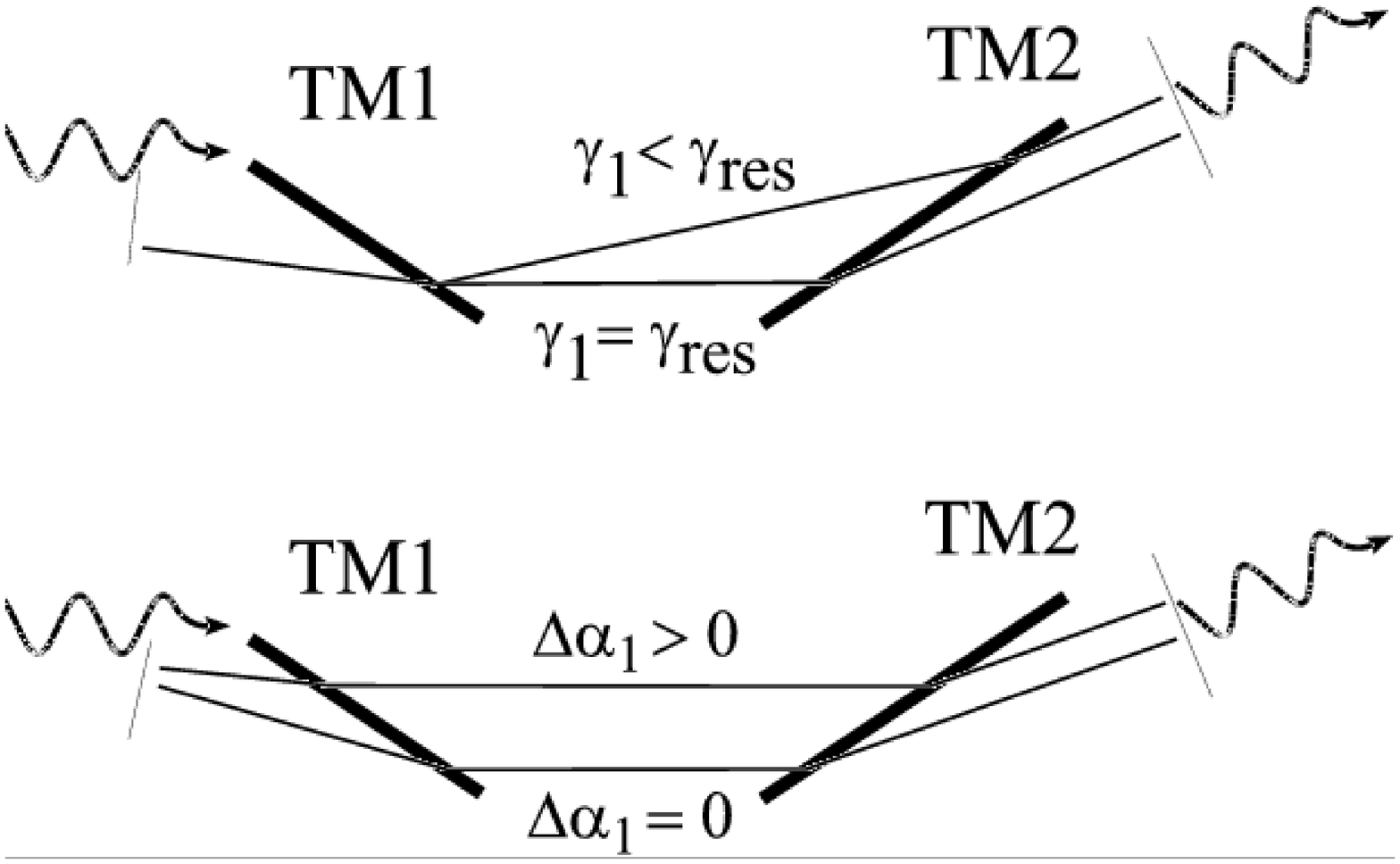} }
{\newline
\small Figure 4. Schema of the phase shift control: negative
dispersion and angular and position sensitivity}

\vskip 12pt


The suggested drift region design shown in Fig.~5 consists of two
magnets. These magnets are assumed to be made of a couple of
magnetic plates of different polarity, with the magnetic field
directed parallel to the $y$-axis. The magnet TM1 serves as an
electron angular and energy analyzer while the magnet TM2 turns
the electron beam further from the original direction, collimates
and directs it to the second wiggler.


\inputEps{220pt}{Figure 5. Drift region geometry.}{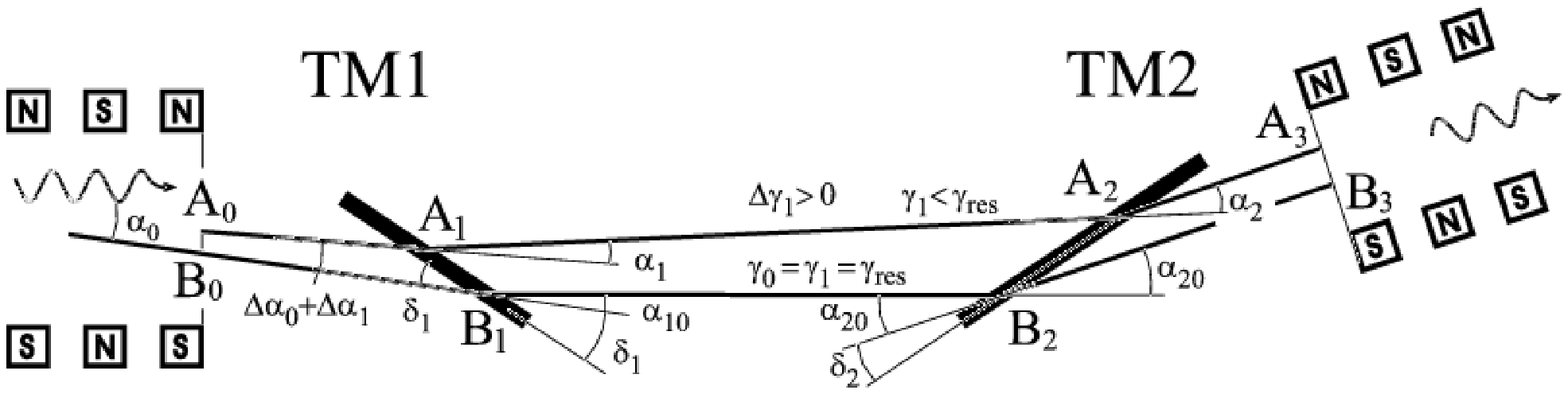}

The drift region geometry is characterized by the lengths
$B_0B_1=L_1$, $B_1B_2=L_2$, $B_2B_3=L_3$ along the reference
trajectory, the angles $\delta_1$, $\delta_2$ between the planes
TM1, TM2 and the reference trajectory intervals $B_0B_1$, $B_2B_3$
respectively. The magnets deflect the electron at angles
$\alpha_1$ and $\alpha_2$, which depend on the electron energy
and, for the second magnet TM2, on the transverse position
$B_2A_2$:

\begin{eqnarray}
\label{alpha-1-alpha-10}
  && \alpha_1 = \alpha_{10} \left(
  1-\displaystyle\frac{\gamma_1-\gamma_{res}}{\gamma_{res}}
  \right),
  \\
\label{alpha-2-alpha-20}
  && \alpha_2 = \alpha_{20} \left(
  1-\displaystyle\frac{\gamma_1-\gamma_{res}}{\gamma_{res}}
  -\displaystyle\frac{B_2A_2}{b} \right),
\end{eqnarray}
where $b$ is the inhomogeneity length for the TM2 magnet. We
assume that the deflection angles $\alpha_{10}$ and $\alpha_{20}$
are large as compared to the initial angular spread of the
electron beam $\Delta\alpha_0$, its field-induced angular spread
$\Delta\alpha_1$, the angles $\theta$ and $\alpha$ between the
wiggler axis and the propagation directions of the light and
electron beams, the angles $\delta_1$ and $\delta_2$ between the
TM1 and TM2 magnets and the corresponding intervals $B_0B_1$ and
$B_2B_3$ of the electron trajectory, and the angular spread
$\alpha_{1}-\alpha_{10}$ and $\alpha_{2}-\alpha_{20}$ introduced
by the magnets TM1 and TM2.

We compare different electron trajectories in the drift region
with the trajectory of a resonant ``reference'' electron, which
has passed the first wiggler along its axis with the unchanged
velocity $\vec v_0=\vec v_1 =\vec v_{res} $ and energy
$\gamma_0=\gamma_1 = \gamma_{res}$. The trajectory $B_0 B_1 B_2
B_3$ is characterized by its transverse positions $B_0$, $B_1$,
$B_2$, $B_3$ at the exit from the first wiggler W1, at the magnets
TM1 and TM2, and at the entrance to the second wiggler W2
respectively.

For an arbitrary electron, which we call ``probe'', the drift
region trajectory $A_0A_1A_2A_3$ is to be found and its length is
to be compared to that of $B_0B_1B_2B_3$. In Fig.~5 the ``probe''
electron was chosen, which increased its energy in the 1-st
wiggler and yet entered the drift region at below-resonant energy
$\gamma_0<\gamma_1<\gamma_{res}$.

The relative phase shift $\Delta\psi_D-\Delta\psi_{D\;ref}$
between the ``probe'' and the ``reference'' electrons is
determined by the path difference $s_A-s_B\equiv
s_{A_0A_1A_2A_3}-s_{B_0B_1B_2B_3}$:

\begin{equation}
\label{Delta-psi-D-Difference}
 \Delta\psi_D-\Delta\psi_{D\;ref} = - \, \frac{\nu_L}{c}
\left( s_{A}-s_{B} \right),
\end{equation}
which is found by adding the path differences at the intervals
where the ``probe'' and ``reference'' electrons move at rather
different angles due to the relatively strong deflection
$\alpha_{10}$ and $\alpha_{20}$   at the magnets TM1 and TM2:
$s_{A}-s_{B}\approx (A_1D_1-C_1B_1)+(D_2A_2-B_2C_2)$, where
$A_1C_1\perp B_0B_1$, $B_1D_1 \perp A_1A_2$, $B_2D_2 \perp
A_1A_2$, and $A_2C_2\perp B_2B_3$, as shown in Fig.~6.

\vskip 12pt

\inputEps{230pt}{Figure 6. Geometry of the drift region path difference.}
{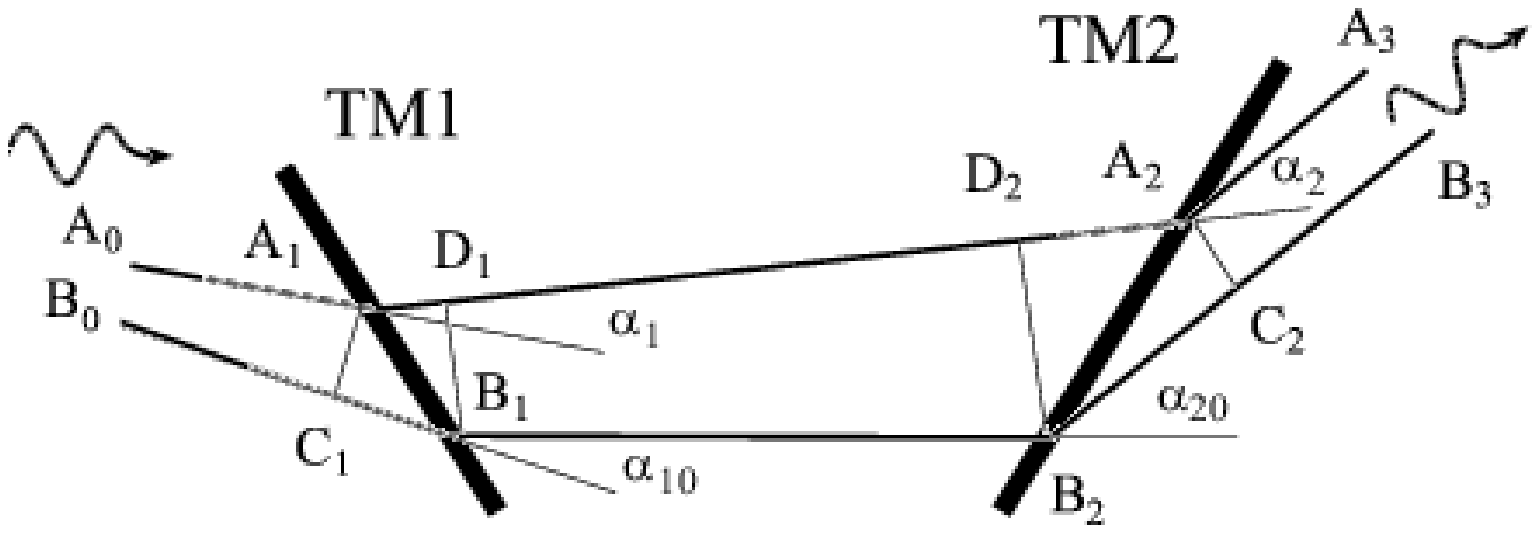}

It is now straightforward to determine the phase shift
$\Delta\psi_D$ of Eq.~(\ref{Delta-psi-D-Difference}). It is
related to that of Eq.~(\ref{Delta-psi-D}) by using
Eqs.~(\ref{Delta-gamma-alpha-x-Delta-Omega-1}),
(\ref{Omega-0-geometry}), (\ref{Delta-Omega-1-geometry}) to
express the parameters $A_0B_0=\Delta x_0+\Delta x_1$,
$\Delta\alpha_1$, $\Delta\alpha_0$, $\Delta\gamma_1$, and
$\Delta\gamma_0$ through the parameters and variables of the
pendulum equation: $\Omega_0$, $\Delta\psi_1$, and
$\Delta\Omega_1=\Omega_1-\Omega_0$. The linearized phase shift
$\Delta\psi_D$ follows:

\begin{eqnarray}
\label{Delta-psi-D-Delta-psi-D-ref-detuning}
 &&
 \Delta\psi_D  {\,}={\,} \Delta\psi_{D\,ref}{\,}+{\,}
 \\ \nonumber && \qquad {\ }+\,
  \Omega_0     \cdot f_{\Omega_0} \,+\,
  \Delta\psi_1 \cdot f_{\Delta\psi_1} \,+\,
  \Delta\Omega_1  \cdot f_{\Delta\Omega_1} \,+\,
  \\ \nonumber  && \qquad {\ }
  +{\ }
  \Delta x_0      \cdot f_{\Delta x_0} \,+\,
  \Delta \alpha_0 \cdot f_{\Delta \alpha_0},
\end{eqnarray}
where the coefficients $f_{\Omega_0}$, $f_{\Delta\Omega_1}$,
$f_{\Delta\psi_1}$, $f_{\Delta x_0}$, and $f_{\Delta \alpha_0}$
describe  the sensitivity of the phase shift to the initial
resonant detuning $\Omega_0$, its wiggler-induced change
$\Delta\Omega_1$, the wiggler-induced change of the slow-phase
$\Delta\psi_1$, the electron beam width $\Delta x_0$ and angular
spread $\Delta\alpha_0$. They are described in the Appendix.

The formulas for the phase shift $\Delta\psi_D$ of
Eqs.~(\ref{Delta-psi-D-Difference}),~(\ref{Delta-psi-D}) match if

 \begin{eqnarray}
 \label{condition-chi-Omega0}
 && f_{\Omega_0} \;=\; \chi \, L_W/c
\\
 \label{condition-chi-DeltaOmega1}
 && f_{\Delta\Omega_1} \;=\; (\chi+\xi) \, L_W/c
\\ \label{Delta-psi-D-f-requirements}
 && f_{\Delta\psi_1} = \vartheta
 \,.
\end{eqnarray}

The drift region dispersion $\chi$ and the sensitivity parameters
$\xi$ and $\vartheta$ of Eq.~(\ref{Delta-psi-D}) can be
implemented by specifying the drift region geometry: the
separation between magnets $L_2=B_1B_2$, the lengths $L_1=B_0B_1$
between the wiggler and the first magnet and the orientation angle
$\delta_1$ of the first magnet:

 \begin{eqnarray}
 L_2&=& -\,\chi\,L_W\,
 \frac{2(1+K^2/2)}{\alpha_{10}\alpha_{20} \gamma_0^2}
 \label{negative-detuning-geometry-condition}
\end{eqnarray}
with the optimum dispersion $ \chi\,=\,-1$,

 \begin{equation}
 \label{drift-region-sensitivity-B0B1-condition}
 L_1 =
 L_W\,
 \frac{\xi+\chi\left(1-{\gamma_0^{-2}(\alpha-\theta)^{-2}}\right)}{\vartheta}
 \, ,
\end{equation}
\begin{equation}
 \label{drift-region-sensitivity-delta1-condition}
  \delta_1 =
  \frac{2\vartheta\,(\theta-\alpha)}{\alpha_{10}(\alpha_{10}+\alpha_{20})}
  \,{\cal K}
  ,
\end{equation}
where the factor
$${\cal K}=
\left[
  \frac{1+K^2/2+\gamma_0^2(\alpha-\theta)^2}
  {\gamma_0^2(\theta-\alpha)^2}
 \right]
 \sim 1
$$

can be dropped for $\gamma_0^2(\theta-\alpha)^2\gg1$. The maximum
angular spread $\Delta\alpha_0$ and the width $\Delta x_0$ of the
electron beam are estimated here for the small deflection angles $
\alpha_{10}$,${\ }\, \alpha_{20}\ll1$. The contributions of the
angular spread $\Delta\alpha_0$ and the transverse beam width
$\Delta x_0$ to the phase shift $\Delta\psi_D$ of
Eq.~(\ref{Delta-psi-D-Delta-psi-D-ref-detuning}) should less than
one. Thus we estimate the maximum $\Delta\alpha_0$ and $ \Delta
x_0$:

 \begin{equation}
 \label{Delta-alpha-0-Estimate}
 \Delta\alpha_0
 \le
 \frac{1}{k_L L_W}
 \;
 \frac{1}{\chi(\theta-\alpha)}
\end{equation}

\begin{equation}
\label{Delta-x-0-Estimate}
 \Delta x_0
 \le
 \frac{1}{k_L}\;
 \frac{1}{\vartheta \;(\theta-\alpha)}\;
 \frac{1}{{\cal K}} .
\end{equation}


The normalized emittance $\varepsilon=\Delta x_0 \,\Delta \alpha_0
\, \gamma_0$ of the electron beam follows:
 \begin{equation}
 \label{emittance-estimate-1}
 \varepsilon \;\le\;
 \frac{1}{\chi\vartheta}\;
 \frac{1}{k_L^2 L_W}\;
 \frac{\gamma_0}{(\alpha-\theta)^2}\;
 \frac{1}{{\cal K}} .
\end{equation}

The limitation on the minimum FEL wavelength $\lambda_L=2\pi
c/\nu_L$ due to the electron beam emittance follows:

\begin{equation}
  \label{lambda-L-emittance-estimate}
 \frac{\lambda_L}{2\pi} \ge \sqrt{
 \frac{L_W \,\varepsilon\, \chi\,\vartheta
 \,(\alpha-\theta)^2}{\gamma_0}
 \; {\cal K}
  }
  \sim
 \sqrt{
 \frac{L_W \,\varepsilon\, \chi\,\vartheta
 }{\gamma_0^3}
  }
\end{equation}

Thus the minimum FELWI wavelength $\lambda_L$ is related to the
electron beam emittannce $\varepsilon$, the drift region
dispersion $\chi$ and the sensitivity parameter $\vartheta$ of
Eq.~(\ref{Delta-psi-D}).

To reduce the electron beam divergence after the drift region, the
inhomogeneity length $b$ is adjusted to collimate the beam. It
divergence after the TM2 magnet due to the energy spread,
$(\alpha_{10}+\alpha_{20})\Delta\gamma_0/\gamma_0$, is compensated
by the focusing term $-\alpha_{20}A_2B_2/b\sim -\alpha_{20}/b\cdot
L_2\alpha_{10}\Delta\gamma_0/\gamma_0$ of
Eq.~(\ref{alpha-2-alpha-20}). Thus the TM2 inhomogeneity length
$b$ is found:

 \begin{equation}
   \label{b-divergence-compensate-estimate}
   b\,=\,L_2\frac{\alpha_{10}\alpha_{20}}{\alpha_{10}+\alpha_{10}}.
 \end{equation}

 We summarize the suggested FEL parameters:
 \newline \noindent
Electron beam:
 \newline \noindent \phantom{.}\quad
 relativistic factor $\gamma_0=15$;
 \newline \noindent \phantom{.}\quad
 velocity angle to the $z$-axis $\alpha=-0.133$\,rad;
 \newline \noindent \phantom{.}\quad
 energy spread $\Delta\gamma_0=1$;
 \newline \noindent \phantom{.}\quad
 emittance $\varepsilon=40 \pi$\,mm$\cdot$mrad;
 \newline \noindent \phantom{.}\quad
 angular spread $\Delta\alpha_0=5\cdot10^{-4}$\,rad;
 \newline \noindent \phantom{.}\quad
 transverse size $\Delta x_0=4$\,mm;
 \newline \noindent
Light beam:
 \newline \noindent \phantom{.}\quad
 wavelength: $\lambda=359$\,$\mu$m;
 \newline \noindent \phantom{.}\quad
 angle to the $z$-axis $\theta=0$;
 \newline \noindent
First wiggler oriented at the angle to the electron and light
beams:
 \newline \noindent \phantom{.}\quad
 period $\lambda_{W1}=2.73$\,cm;
 \newline \noindent \phantom{.}\quad
 number of periods $N_{W1}=36$;
 \newline \noindent \phantom{.}\quad
 length $L_{W1}=0.98$\,m;
 \newline \noindent \phantom{.}\quad
 parameter $K=1.27$;
 \newline \noindent
Second wiggler oriented along the electron and light beams:
 \newline \noindent \phantom{.}\quad
 period $\lambda_{W2}=8.79$\,cm;
 \newline \noindent \phantom{.}\quad
 number of periods $N_{W2}=36$;
 \newline \noindent \phantom{.}\quad
 length $L_{W2}=3.16$\,m;
 \newline \noindent \phantom{.}\quad
 parameter $K=1.27$;
 \newline \noindent
Drift region:
 \newline \noindent \phantom{.}\quad
 deflection angle at TM1 $\alpha_{10}=0.45$\,rad;
 \newline \noindent \phantom{.}\quad
 orientation of magnets
 \newline \noindent \phantom{.}\quad
 $\delta_1=\delta_2=3\alpha_{10}\Delta\gamma_0
 /\gamma_0=0.09$\,rad;
 \newline \noindent \phantom{.}\quad
 deflection angle at TM2
 \newline \noindent \phantom{.}\quad
 $\alpha_{20}=3\delta_1=3\delta_2=0.27$\,rad;
 \newline \noindent \phantom{.}\quad
 TM2 inhomogeneity length $b=12.7$\,cm;
 \newline \noindent \phantom{.}\quad
 distances between magnets: $L_1=L_2=0.1$\,m.

The suggested parameters provide the phase shift $\Delta\psi_D$
suitable for free-electron lasing without inversion:

 \begin{eqnarray}
 \nonumber
 \Delta\psi_D&=&
 const -\Omega_1 \,T  \,+\, 0.6 \Delta\Omega_1\,T
\,-\, 0.07 \Delta\psi_1
\\
& & -\,2\cdot 10^3 \Delta\alpha_0 + 250\Delta x_0
\end{eqnarray}

Thus the free-electron lasing without inversion can be achieved in
the submillimeter domain with the available electron beams.

\section{The FELWI gain}

Once the drift region for the FELWI is described, we proceed with
the calculations of the gain for the phase shift $\Delta\psi_D$ of
Eq.~(\ref{Delta-psi-D}). The change of the vector potential of the
amplified wave, Eq.~(\ref{Delta-AL-two-integrals}), is sensitive
to the drift region phase shift $\Delta \psi_D$. A particular form
of the phase shift

\begin{equation}
\label{Delta-psi-D0}
 \Delta\psi_D^{(0)}
  = \varsigma + \chi \,\Omega_0 \, T
  = \varsigma + \chi \,\Omega_1 \, T - \chi \,\Delta\Omega_1 \, T .
\end{equation}

was analyzed \cite{6} and it was shown that such a phase shift
indeed provides an FELWI. In this paper, we extend the results of
\cite{6} and find the gain for an arbitrary linear phase shift
$\Delta\psi_D$ of Eq.~(\ref{Delta-psi-D}). The gain is found by
calculating the integrals of Eq.~(\ref{Delta-AL-two-integrals})
over the lengths of both wigglers. The contribution of the first
wiggler is described by the first integral of
Eq.~(\ref{Delta-AL-two-integrals}), which leads to the usual
expression for the gain, $G={ \Delta |A_L|^2 }/{|A_L|^2}$, of a
one-wiggler FEL (see \cite{Scully-1976} and Appendix). The
two-wiggler FELWI gain is found by using the perturbation theory
approach and considering the $\Delta\psi_D^{(0)}$ part of the
phase shift of Eq.~(\ref{Delta-psi-D0}) as the zero-order
approximation to $\Delta\psi_D$ of Eq.~(\ref{Delta-psi-D}). The
wiggler-induced correction
$\Delta\psi_D-\Delta\psi_D^{(0)}=(\xi+\chi) \, \Delta \Omega_1 \,
T + \vartheta \Delta\psi_1$ is treated as a perturbation.
Expanding the right-hand side of
Eq.~(\ref{Delta-AL-two-integrals}) over the first-order terms in
the electron-light interaction, we find the change of the laser
vector potential $A_L$ and the FELWI gain:

%
%

\begin{eqnarray}
\label{Gain-FELWI-Book} &&
G (u)=
\frac{e^4 H_W^2 n_e  L^3 }
     {2 m^3 \gamma^3 c^4  \beta_{z\,0}^2 k_W \varepsilon_0}
\\ \nonumber
&& \, \Biggl< \,-\,
  \frac{\partial }{\partial u} \,
\frac{\sin^2{u}\,\cos^2(u+\frac{1}{2}\Delta\psi_D^{(0)}) }{u^2}
 \\
 && \nonumber
 +\, (\xi+\chi) \frac{\sin^2 u \,
\sin(2u+\Delta\psi_D^{(0)})}{u^2}
\\
&& \nonumber
 -\,
  \frac{\vartheta}{4}\,
  \frac{\partial}{\partial u}\,
  \frac{\sin^2 u\,\cos(2u+\Delta\psi_D)}{u^2}
  \Biggr>_{\Delta\psi_D^{(0)}=\varsigma + \chi \, \Omega_0 \, T}
\end{eqnarray}
where $u=\Omega_0 T/2=(\vec k_L+\vec k_W)\cdot(\vec v-\vec
v_{res})\,T/2$. The gain of Eq.~(\ref{Gain-FELWI-Book}) is
expressed as a sum of the three terms. The first term has a form
of the total derivative and is similar to the known expression
\cite{Datolli-PRE-V61-p7052} for the optical klystron gain. The
other two terms result from the first-order corrections to the
drift region phase shift $\Delta\psi_D=\varsigma+\chi \Omega_1 T
+\xi \Delta\Omega_1 + \vartheta \Delta\psi_1$. For a negative
drift region dispersion, e.g., $\chi=-1$, it is possible to
achieve free-electron lasing without inversion by changing the
sensitivity parameters $\xi$ or $\vartheta$, as it is shown in
Fig.~7. The four curves shown correspond a
 usual FEL gain profile $G(u)$ for $\xi=0$ and
 $\vartheta=0$ (curve a) and
 FELWI gain profiles $G(U)$, for increasing
 sensitivities of the drift region phase shift $\delta\psi_D$
 to $\xi$ and $\vartheta$:
 b) $\xi=0.5$, $\vartheta=0$ or $\xi=0$, $\vartheta=1$;
 c) $\xi=1$, $\vartheta=0$ or $\xi=0$, $\vartheta=2$;
 d) $\xi=1.5$, $\vartheta=0$ or $\xi=0$, $\vartheta=3$.
 For other drift region dispersion $\chi\ne-1$ the sensitivities $\xi$ and
 $\vartheta$ change the shape of the gain profile $G(u)$ differently.

\vskip 12pt

\inputEps{120pt}{Figure~7.
The gain profile $G(u)$ for different sensitivities of the drift
region  phase shift $\Delta\psi_D$ to the field-induced change of
the resonant detuning $\Delta\Omega_1$ and the field-induced
change of the slow phase $\Delta\psi_1$. The curves a) --- d) are
discribed in the text.}{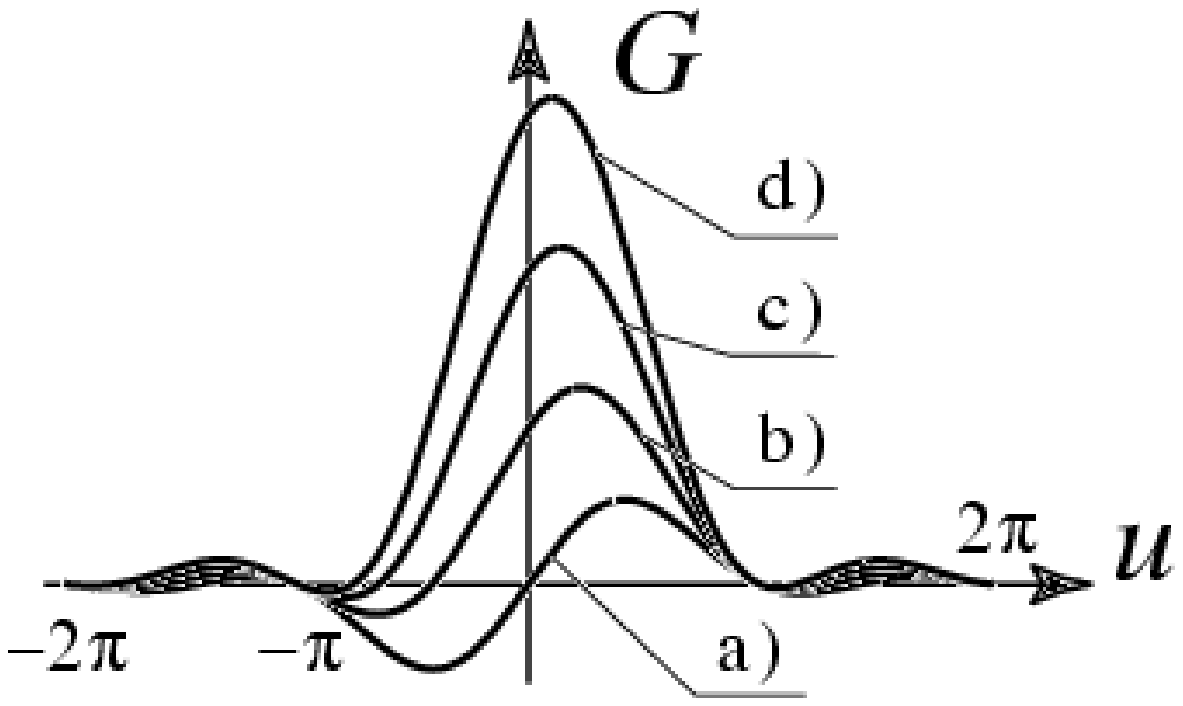}


To determine the hot-beam gain we average the FELWI gain of
Eq.~(\ref{Gain-FELWI-Book}) over the broad electron distribution
function $f({\gamma})$, such that $\Delta \gamma_0 /\gamma_0 \gg
1/N_W={2\pi}/{L_W k_W}$. The hot-beam FELWI gain $G_{hot}=\int
G(u)f(\gamma)d u$ follows:

\newcommand{\sign}{{\mathrm{sign}}}

\begin{eqnarray}
\label{Gain-FELWI-Hot} G_{hot} &=& \frac{e^4 H_W^2 n_e L^3
f({\gamma}_{res})}
     {m^3 \gamma^3 c^4 \beta_{z\,0}^2 k_W \varepsilon_0}
\, \frac{\pi}{4} \, \sin \varsigma \, \cdot
\\
\nonumber &&\Bigl(\xi\cdot I_{\xi} \;+\; \vartheta\cdot
I_{\vartheta} \Bigr)
\end{eqnarray}
where
\begin{eqnarray}
I_{\xi} &=&\Bigl( |\chi|+ |\chi+2| - 2 |\chi+1| \Bigr)
\\
\nonumber
 I_{\vartheta} &=&-\frac{1}{2}\, \Bigl( 2\ |1 + \chi| -
 2\ |2 + \chi| + \chi^2\ \sign[\chi]
\\  &&
 - 2\ \sign[1 + \chi] -
      4\ \chi\ \sign[1 + \chi]
\\ \nonumber &&
       - 2\ \chi^2\ \sign[1 + \chi] + 4\ \sign[2 + \chi]
\\ \nonumber &&
        +
      4\ \chi\ \sign[2 + \chi] + \chi^2\ \sign[2 + \chi]
      \Bigr)
\end{eqnarray}

The hot-beam FELWI gain $G_{hot}$ of Eq.~(\ref{Gain-FELWI-Hot}) is
proportional to the sensitivity $\xi$ and $\vartheta$ of the drift
region to the field-induced change of the resonant detuning and
the slow phase in the first wiggler. The gain is non-zero only for
the negative dispersion $-2<\chi<0$ and achieves maximum at
$\chi=-1$, as shown in Fig.~8. Thus the combination of the proper
negative dispersion and the sensitivity of the phase shift
$\Delta\psi_D$ to $\Delta\Omega_1$ and $\Delta\psi_1$ provides
lasing without inversion in the two-wiggler system: the gain
$G(u)$ is enhanced and its shape changes from an odd to a unipolar
profile. For a usual FEL or an optical klystron with positive
dispersion, $\chi>0$, the hot-beam gain turns to zero, because in
this case the gain $G(u)$ has positive and negative peaks, which
compensate each other. Note that the dependence of the phase shift
$\Delta\psi_D$ on the field-induced change of the resonant
detuning and phase change in the first wiggler can not be
neglected, even though the contribution of these terms,
$(\Omega_1-\Omega_0)T$ and $\psi_1-\varphi_0$, can be small as
compared to $|\Omega_1|T$: without these terms the hot-beam FELWI
gain turns to zero according to the Madey's theorem
\cite{5},~\cite{Madey}.

\inputEps{180pt}{Figure~8.
The hot-beam gain of a FELWI determined by the integrals $I_{\xi}$
and $I_{\vartheta}$ as functions of the drift region dispersion
$\chi$.
}{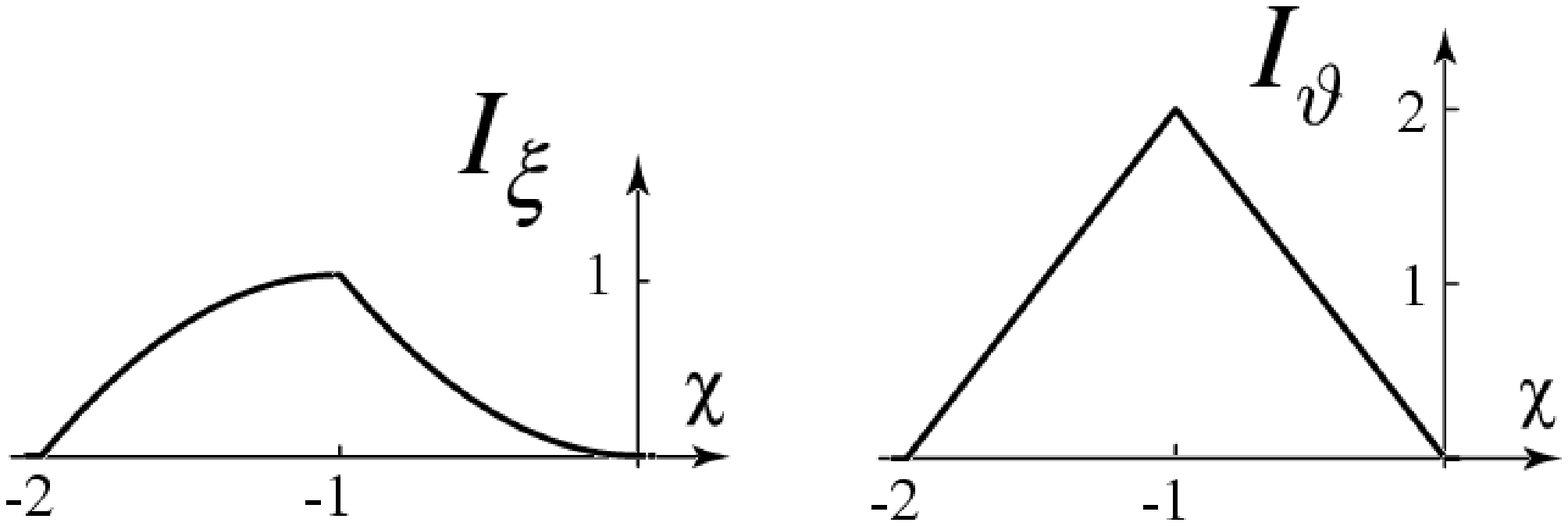}

\subsection{Phase distribution in the FELWI}

To illustrate the physics of the FELWI operation, we present in
Fig.~9 the phase space distribution of the electron beam while it
passes the FELWI. Depending on the initial phase $\phi_0$, both
the electron energy $\gamma$ and its dimensionless resonant
detuning $u=\Omega T$ are increased or decreased, as it is shown
in Fig.~9b. The phase shift $\Delta\psi_D$ introduced by the drift
region rearranges electrons in the phase space according to their
resonant detuning and its change in the first wiggler, Fig.~9b.
Then in the second wiggler most electrons loose their energy and
contribute to the light amplification. The resulting gain is
positive for any resonant detuning, either positive or negative,
as it is shown in Fig.~9d. The unipolar gain profile is a
characteristic features of a FELWI.

\vskip 12pt


\inputEps{220pt}{Figure~9. Phase space ditribution of the electron beam at different
part of the FELWI:
\newline\noindent{}
a)~initial distribution;
\newline\noindent{}
b)~distribution and gain after the first wiggler;
\newline\noindent{}
c)~distribution after the drift region;
\newline\noindent{}
d)~distribution and gain after the second wiggler.
}{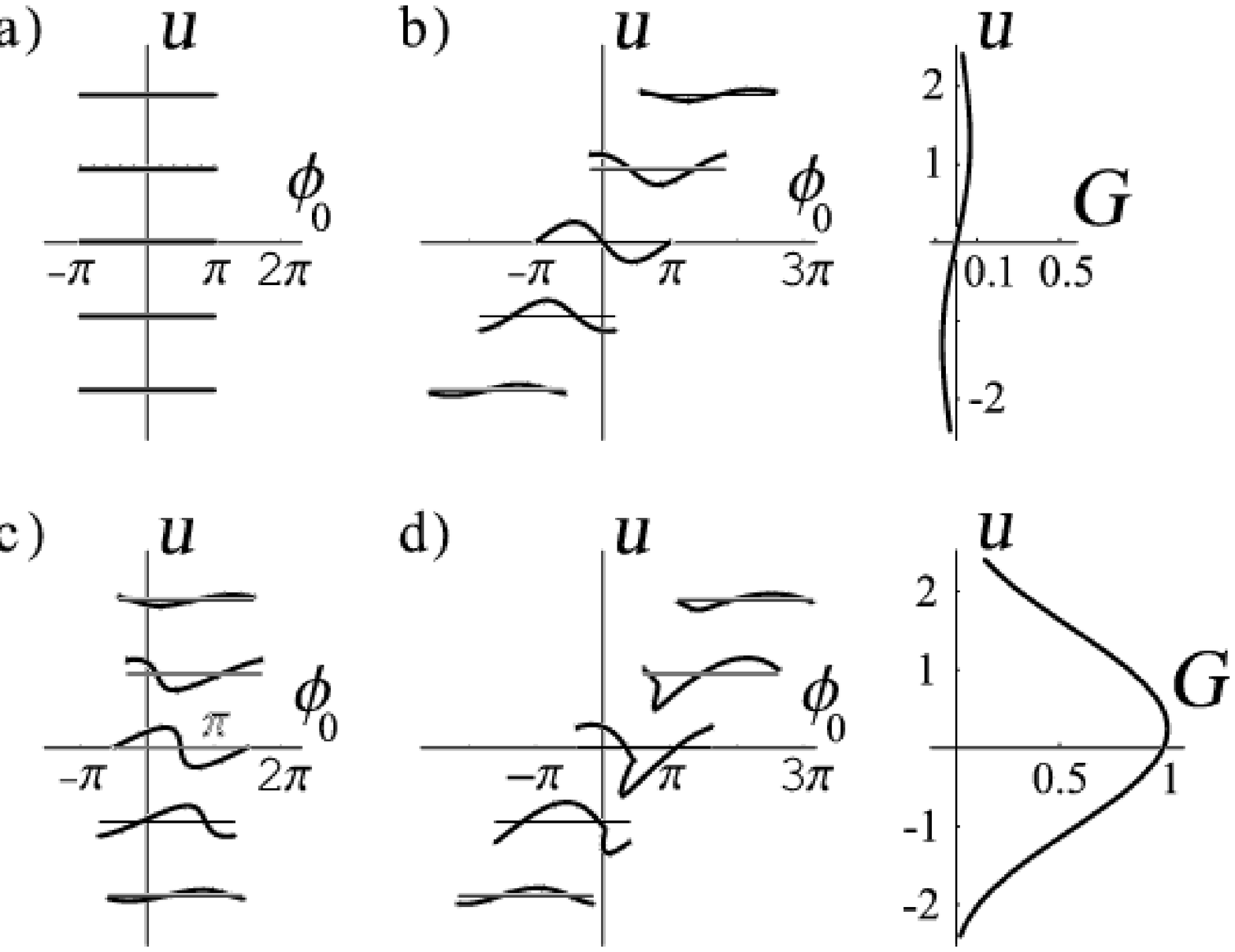}

\section{Conclusions}

A realistic design of the drift region, which implements the FELWI
gain is suggested and analyzed in detail for the first time. The
design simplicity is achieved due to the overall turning of the
electron beam in the drift region. The geometry of the drift
region is related to the parameters describing the FELWI gain: the
drift region dispersion and sensitivity to the field-induced
changes of the resonant detuning and phase. The gain profile and
the hot-beam gain of an FELWI is found for a linear phase shift of
the drift region. It is shown that free-electron lasing without
inversion requires the negative drift region dispersion, which
should be limited in the certain range. Another requirement to the
phase shift is that the electron path length through the drift
region should be sensitive to the direction of the electron
velocity or the transverse electron position at the exit from the
first wiggler. The minimum FELWI wavelength is found and related
to the drift region sensitivity parameters, to the transverse size
and angular spread of the electron beam. For available electron
beams the suggested FELWI can operate in the submillimeter domain.

The described method for the phase control, which is applied here
for an optical klystron, can be adjusted, in principle, for other
applications. It can be used for phase corrections and gain
enhancement in storage ring FELs because storage rings achieve
isochronous motion of the electrons by insertion negative
dispersion sections \cite{ring-megative-dispersion}, which can
serve as a key elements of a drift region for a storage ring
FELWI.

\section{Acknowledgements}

The authors are grateful to L.~Narducci, D.~Mills, 
B.~Adams, and C.~Keitel for illuminating discussions. 
The authors acknowledge the support from the 
Office of Naval Research, 
the National Science Foundation, 
Texas Engineering Experiment Station, 
the Robert~A. Welch Foundation, and TATP.
G.K., M.S., Y.R., and A.I.A. 
acknowledge the support of 
the US-Israel BSF, 
as well as the Russian Foundation for Basic Research
grant 02-02-17135 (A.I.A.).

\appendix
\section{Equations of motion and gain}

The text is splitted into subsections for editing purposes.

\subsection{Hamiltonian}

The classical electron dynamics in an FEL is described by
Hamiltonian

\begin{equation}
\label{hamiltonian-MKS-appendix}
 H \equiv \gamma m c^2 = \sqrt{
(\vec{p}-e\vec{A} )^2 c^2 + m^2 c^4}
\end{equation}
where $c$ is the speed of light, $e$, $m$, $\gamma$, and $\vec p$
are the electron charge, mass, Lorentz factor, and canonical
momentum,

\begin{equation}
\label{vector-potential-appendix} \vec{A} \equiv \vec{e}_y A_y =
\vec{e}_y ( A_L   e^{i\psi_L} + A_W   e^{i\psi_W} + \; C.C. )
\end{equation}

is the vector potential of the combined electromagnetic field of
the laser (designated by a subscript $L$) and the wiggler
(designated by a subscript $W$). The electron and laser beams
propagate at the small angles $\alpha$ and $\theta$ to the axis of
the wiggler, as in Fig.~2. The $z$-axis is directed along the
wiggler, so that $\psi_W=k_W z$, and the phase of the laser field
$\psi_L$ equals $k_L ( z \cos\theta + x \sin \theta)-\nu_L t +
\phi_0$.

%

\subsection{Dynamics along {\boldmath$y$}-axis}

The Hamiltonian (\ref{hamiltonian-MKS-appendix}) does not depend
on $y$. Therefore the electron canonical momentum in the
$y$-direction is unchanged and is assumed to be zero:

\begin{eqnarray}
\label{hamilt-eqs-dpydt-appendix} && \!\!\!\!\!\! \frac{dp_y}{dt}
= - \frac{\partial H}{\partial y} = 0, \quad p_y(0)=p_y(t)=0,
\\
\label{hamilt-eqs-dydt-appendix} && \!\!\!\!\!\! \frac{dy}{dt} =
\left. \frac{\partial H}{\partial p_y} \right|_{p_y=0} = \left.
\frac{p_y - e A_y}{\gamma m} \right|_{p_y=0}= -\, \frac{  e
A_y}{\gamma m}.
\end{eqnarray}


\subsection{Zero-order electron dynamics}

The electron interaction with the laser field $A_L$ is treated as
a perturbation \cite{Brau}. In the absence of the laser field the
magnetic field of the wiggler $A_y=A_W e^{i\psi_W}+C.C.$ cause
oscillations (\ref{hamilt-eqs-dydt-appendix}) of the electron
velocity $\vec\beta=\vec v /c$ and reduces its projection to the
$z$-axis:

\begin{eqnarray}
\label{bety-wiggler-field-appendix} \beta_y {\ } \equiv {\
}\frac{v_y}{c} & =&-\, \frac{K \sin(k_W z)}{\gamma },
\\
\frac{d p_x}{dt} &=& -\frac{\partial H}{\partial x} {\ }= {\ }0
{\quad}p_x{\ }={\ }const
\\
\label{betx-wiggler-field-appendix}
 \beta_x &=& \frac{\ \ \partial H}{c\
 \partial p_x} {\ }= {\ } \frac{p_x}{mc\gamma} {\ }= {\ } \alpha
\\
\label{betz-wiggler-field-appendix}
 \beta_{z} &=&
 \sqrt{1-\frac{1}{\gamma^2}-\beta_x^2-\beta_y^2}
\\ \nonumber
& = & 1-\frac{1+\alpha^2\gamma^2+K^2 \,\sin^2(k_W z)}{2 \gamma^2}
\\ \nonumber
\langle \beta_{z} \rangle
& = & 1-\frac{1+\alpha^2\gamma^2+K^2
/2}{2 \gamma^2} ,
\end{eqnarray}
where angular brackets denote time averaging, and $K$ is the
wiggler parameter
\begin{equation}
K= \frac{2 e A_W }{ m c} = \frac{e H_W}{m c k_W}.
\end{equation}

\subsection{First-order electron dynamics}

The first-order corrections to the electron canonical momentum
$\vec p$, energy $mc^2\gamma$, and coordinates $x$, $z$ are found
by the perturbation theory over the electron-light interaction.
Eqs.~(\ref{bety-wiggler-field-appendix})-(\ref{betz-wiggler-field-appendix})
provide the zero-order approximation and account for the $(A_W
e^{i\psi_W}+C.C.)^2$ part of the squared vector potential $A_y$ of
Eq.~(\ref{vector-potential-appendix}). The resonant approximation
reduces the remaining part of the squared vector potential $A_y^2
- A_W^2$ to the ponderomotive potential $2 (A_L A_W e^{i \psi}
+\; C.C.)$ with the slow phase

\begin{equation}
\label{psi-definition-appendix} \psi = \psi_L+\psi_W = (\vec
k_L+\vec k_W)\cdot \vec r - \nu_L t + \phi_0.
\end{equation}

The condition $\dot\psi=0$ determines the resonant frequency
$\nu_{res}$ of the FEL:

\begin{equation}
\label{nu-Res-appendix} \nu_{res}=\frac{2 c k_W
\gamma^2}{1+K^2/2+\gamma^2(\alpha-\theta)^2}.
\end{equation}

The equations for the electron motion expressed in terms of the
slow phase $\psi$ follow:

\begin{eqnarray}
\label{dpxdtA2-appendix} \frac{dp_x}{dt} &=& \frac{e^2}{\gamma m}
2 A_W A_L k_L \sin \theta \, \sin\psi
\\
\label{dpzdtA2-appendix} \frac{dp_z}{dt} &=& \frac{e^2}{\gamma m}
2 A_W A_L (k_L \cos\theta + k_W) \, \sin\psi
\\
\label{dgammadtA2-appendix}  \frac{d\gamma}{dt} &=&
\frac{e^2}{\gamma m^2 c^2} 2 A_W A_L \, \nu_L  \, \sin\psi.
\\
\label{dxdt1-appendix} \displaystyle{ \frac{d^2x}{dt^2} } &=&
\frac{d}{dt} \left( \frac{p_x}{\gamma m} \right) \; = \;
\frac{e^2}{m^2\gamma^2} 2 A_W A_L \cdot
\\ \nonumber
&  & {\ }
 \cdot
 \left[ k_L \sin\theta - \nu_L/c\cdot \beta_{x}
 \right] \sin\psi
\\
\label{dzdt1-appendix} \displaystyle{ \frac{d^2z}{dt^2} } &=&
\frac{d}{dt} \left( \frac{p_z}{\gamma m} \right) \; = \;
\frac{e^2}{m^2\gamma^2} 2 A_W A_L \cdot
\\
\nonumber & & {\ } \cdot \left[ k_L \cos\theta + k_W -
\nu_L/c\cdot \beta_{z} \right] \sin\psi,
\end{eqnarray}

The electron dynamics is governed by the $\sin \psi$ factors in
the right-hand sides of
Eqs.~(\ref{dpxdtA2-appendix})-(\ref{dzdt1-appendix}). Thus it is
reduced to the evolution of the slow phase $\psi$, which is
determined by
Eqs.~(\ref{pendulum-equation-system-dpsidt})--(\ref{pendulum-equation-system-dOmegadt})
below. This reduction leads to the correlation between the
field-induced deflection angle $\Delta\alpha_1$, the changes of
the electron energy $\Delta\gamma_1$ and its velocity components
$\Delta v_{x1}$, $\Delta v_{z1}$:

\begin{eqnarray}
\label{Delta-alpha-Delta-gamma-appendix}
  \Delta\alpha_1 &=& \frac{\Delta v_{x1}}{c} {\ }={\ }
  \Delta\gamma_1
  \,\frac{\theta-\alpha}{\gamma},
\\
\label{Delta-vx-Delta-vz-appendix}
  \Delta v_{x1} &=& \Delta v_{z1}\,
  \frac{\theta-\alpha}{\displaystyle{\frac{1+K^2/2}{\gamma^2}}+\alpha(\alpha-\theta)}.
\end{eqnarray}

We interpret this correlation in terms of the photon emission and
absorption in the wiggler: the electrons absorbing photons get
the momentum kick towards the light beam.

\subsection{Pendulum equation}

The evolution of the slow phase $\psi$ of
Eq.~(\ref{psi-definition-appendix}) is determined by the pendulum
equation~(\ref{pendulum-equation-system-dpsidt}),~(\ref{pendulum-equation-system-dOmegadt}),
which follows from
Eqs.~(\ref{dpxdtA2-appendix})--(\ref{dzdt1-appendix}).

\subsection{Zero- and first-order approximation to the pendulum equation}

The pendulum equation
(\ref{pendulum-equation-system-dpsidt})--(\ref{Omega0-definition})
is solved by considering the electron-light interaction as a
perturbation, $\psi_I(t)=\psi^{(0)}_{I}(t)+\psi^{(1)}_{I}(t)$ in
the first wiggler, $0<t<T$, and
$\psi_{II}(t)=\psi^{(0)}_{II}(t)+\psi^{(1)}_{II}(t)+ \Delta\psi_D$
in the second wiggler, $T<t<2T$. The zero-order approximation to
the phase $\psi$ is linear in time:
$\psi^{(0)}_I=\phi_0+\Omega_{0}t$ and $\psi^{(0)}_{II}=\phi_0+
\Delta\psi_D +\Omega_0 t$ in the first and second wigglers
respectively. The first-order corrections $\psi^{(1)}_I$ and
$\psi^{(1)}_{II}$ are found by the procedure described in
\cite{4},~\cite{5}.

To relate the pendulum equation in the wiggler with the initial
conditions for the electron beam and with the electron dynamics in
the drift region, we need to express the initial resonant detuning
$\Omega_0$ of Eq.~(\ref{Omega0-definition}) and its change in the
first wiggler $\Delta\Omega_1=\Omega_1-\Omega_0$ through the
electron beam parameters: the initial angular deviation
$\Delta\alpha_0$ of the electron velocity from the beam
propagation direction $\alpha$, the initial deviation of the
electron relativistic factor $\gamma$ from the resonance,
$\Delta\gamma_0=\gamma_0-\gamma_{res}$,  and the wiggler-induced
change of the relativistic factor
$\Delta\gamma_1=\gamma_1-\gamma_0$:

\begin{eqnarray}
 \label{Omega-0-geometry}
 \Omega_0&=&
 \Delta\gamma_0 \,
 \frac{2 c k_W (1+K^2/2)}{\gamma (1+K^2/2+\gamma^2(\alpha-\theta)^2)}
 \\ \nonumber
 &+&
 \Delta\alpha_0 \,
 \frac{2c k_W \gamma^2 (\theta-\alpha)}{\gamma(1+K^2/2+\gamma^2(\alpha-\theta)^2)}
 \\ \nonumber
 &=&
  \Delta\gamma_0 \,\frac{\nu_L(1+K^2/2)}{\gamma^3}
  +
  \Delta\alpha_0 \, \nu_L\,(\theta-\alpha)
  \\ \nonumber
  \Delta\Omega_1 &=&
  \Delta\alpha_1\, \frac{\nu_L(1+K^2/2+\gamma^2(\alpha-\theta)^2)}{\gamma^2(\theta-\alpha)}
  \\ \label{Delta-Omega-1-geometry}
  &=& \Delta\alpha_1\,\frac{2 c k_W}{\theta-\alpha} {\ }={\ }
  \Delta\gamma_1\,\frac{2 c k_W}{\gamma}.
\end{eqnarray}

\subsection{Field evolution - 1}

The field evolution in the FELWI is described by
Eqs.~(\ref{AL-Slow-Envelope})-(\ref{Delta-AL-two-integrals}).
\subsection{One-wiggler gain}

The contribution of the first wiggler is described by the first
integral of Eq.~(\ref{Delta-AL-two-integrals}), which leads to the
usual expression \cite{Scully-1976} for the gain, $G={ \Delta
|A_L|^2 }/{|A_L|^2}$, of a one-wiggler FEL:

\begin{equation}
\label{Gain-1-wiggler} G  =  -\, \frac{c^2  e^4  H_W^2 \, n_e
L_W^3}
     {8 \varepsilon_0 \,(m c^2 \gamma)^3 k_W }
\, \left<  \frac{d}{du} \, \frac{\sin^2 u}{u^2}
 \right>_{u = \Omega_0 L_W / 2 c},
\end{equation}

where $\varepsilon_0=8.85\cdot10^{-12} F/m$ is the permittivity of
vacuum (SI units), $H_W$ is the amplitude of the magnetic field of
the wiggler, $L_W$ is the wiggler length, and $n_e$ is the
concentration of electrons. This is a well-known result for the
odd gain profile of an FEL: for energies above the resonance most
of electrons contribute to the amplification, while for
below-resonant energies, most of electrons contribute to the
absorption of radiation. For all energies, below or above the
resonance, there are electrons contributing to the amplification
or absorption of radiation, depending on the phase of the laser
field at the moment of the electron entry to the wiggler.

\section{Drift region geometry}

The path differences acquired at each magnet are found from the
geometry of the quadrangles $A_1C_1B_1D_1$ and $A_2C_2B_2D_2$ of
Fig.~6:

\begin{eqnarray}
\nonumber
 A_1D_1-C_1B_1{\ } = {\ }A_1B_1
 (\cos(\alpha_{10}+\delta_1)-\cos\delta_1),
\\
 A_2D_2-C_2B_2 {\ } = {\ }A_2B_2
 (\cos(\alpha_{20}+\delta_2)-\cos\delta_2).
\end{eqnarray}
The contributions of other trajectory intervals to the path
difference is of the second order of the small angles
$\Delta\alpha_0$, $\Delta\alpha_1$, $\alpha$, $\theta$,
$\delta_1$, $\delta_2$, $\alpha_1-\alpha_{10}$, and ,
$\alpha_2-\alpha_{20}$ and thus can be neglected. The intervals
$A_1B_1$ and $A_2B_2$ follow:

\begin{eqnarray}
 \nonumber
 A_1B_1 & = & ( A_0B_0 + B_0B_1
 (\Delta\alpha_0+\Delta\alpha_1) \sin\delta_1,
\\ \nonumber
 A_2B_2 & = &
 B_1B_2 \cdot
 \frac{\alpha_1-\alpha_0+\Delta\alpha_0+\Delta\alpha_1}{\sin(\alpha_{20}+\delta_2)}
 +
\\
&& \!\!\!\!\!\! + {\ } A_1B_1 \cdot
\frac{\sin(\delta_1+\alpha_{10})}{\sin(\delta_2+\alpha_{20})}
\end{eqnarray}

yielding the deviation of the phase shift $\Delta\psi_D$ of the
``probe'' electron relative to the phase shift
$\Delta\psi_{D\,ref}$ of the ``reference'' electron:

\begin{eqnarray}
\label{Delta-psi-D-Delta-psi-D-ref-geometry-appendix}
 &&
 \Delta\psi_D  {\,}-{\,} \Delta\psi_{D\,ref}{\,}={\,}
  \\ \nonumber  &&
 -\,\frac{\nu_L}{c} \,\Bigl[
B_1B_2
\bigl[\bigl((\alpha_1-\alpha_{10})+(\Delta\alpha_0+\Delta\alpha_1)\bigr)
\cdot
 \\ \nonumber  && \qquad
 (\cot(\alpha_{20} + \delta_2) - \cos \delta_2 \csc(\alpha_{20} + \delta_2))
\bigl]{\ }+{\ }
 \\ \nonumber  &&
 {\ }+{\ }
 \bigl(A_0B_0 + B_0B_1 (\Delta\alpha_0+\Delta\alpha_1)
 \bigr)
 \cdot
 \\ \nonumber &&  \qquad
 \bigl[
   (\cos(\alpha_{10} + \delta_1)-\cos \delta_1){\ }+
 \\ \nonumber && \qquad {\ }
   (\cot(\alpha_{20} + \delta_2) -
    \cos\delta_2 \csc(\alpha_{20} + \delta_2)
    )\cdot
 \\ \nonumber && \qquad {\ }
    \sin(\alpha_{10} + \delta_1)
 \bigr]
 \,\Bigr]
\end{eqnarray}

with $\alpha_{1}-\alpha_{10}$ and $\alpha_{2}-\alpha_{20}$ given
by Eqs.~(\ref{alpha-1-alpha-10}),~(\ref{alpha-2-alpha-20}). The
drift region geometry can be chosen so that the linearized phase
shift of Eq.~(\ref{Delta-psi-D-Delta-psi-D-ref-geometry-appendix})
coincides with that of Eq.~(\ref{Delta-psi-D}). To find the
corresponding requirements, the electron position $A_0B_0=\Delta
x_0+\Delta x_1$ at the exit from the first wiggler, the angular
and energy deviations $\Delta\alpha_1$, $\Delta\alpha_0$,
$\Delta\gamma_1$, and $\Delta\gamma_0$ are related to the
field-induced phase change $\Delta\psi_1$, the initial resonant
detuning $\Omega_0$ and its change in the first wiggler
$\Delta\Omega_1=\Omega_1-\Omega_0$ given by
Eqs.~(\ref{Delta-gamma-alpha-x-Delta-Omega-1}),
(\ref{Omega-0-geometry}), (\ref{Delta-Omega-1-geometry}). The
linearized drift region phase shift $\Delta\psi_D$ follows:

\begin{eqnarray}
\label{Delta-psi-D-Delta-psi-D-ref-detuning-appendix}
 &&
 \Delta\psi_D  {\,}={\,} \Delta\psi_{D\,ref}{\,}+{\,}
 \\ \nonumber && \qquad {\ }+\,
  \Omega_0     \cdot f_{\Omega_0} \,+\,
  \Delta\psi_1 \cdot f_{\Delta\psi_1} \,+\,
  \Delta\Omega_1  \cdot f_{\Delta\Omega_1} \,+\,
  \\ \nonumber  && \qquad {\ }
  +{\ }
  \Delta x_0      \cdot f_{\Delta x_0} \,+\,
  \Delta \alpha_0 \cdot f_{\Delta \alpha_0},
\end{eqnarray}

where the coefficients $f_{\Omega_0}$, $f_{\Delta\Omega_1}$,
$f_{\Delta\psi_1}$, $f_{\Delta x_0}$, and $f_{\Delta \alpha_0}$
describe  the sensitivity of the phase shift to the initial
resonant detuning $\Omega_0$, its wiggler-induced change
$\Delta\Omega_1$, the wiggler-induced change of the slow-phase
$\Delta\psi_1$, the electron beam width $\Delta x_0$ and angular
spread $\Delta\alpha_0$:

\begin{eqnarray}
\nonumber
  f_{\Omega_0} &=&
  -\,\frac{\alpha_{10}\tan(\alpha_{20}/2)\gamma_0^2}{c(1+K^2/2)}\,B_1B_2
  \qquad
  \\  &\approx &
  -\,\frac{\alpha_{10}\alpha_{20}\gamma_0^2}{2c(1+K^2/2)}\,B_1B_2
\end{eqnarray}
\begin{eqnarray}
\nonumber \!\!\!\!\!\!\!\!\!\!\!\!
  f_{\Delta\psi_1} =\;
  \frac{\delta_1 \gamma_0^2 (\alpha - \theta)
        (\cos\alpha_{10}-1-\sin\alpha_{10}\tan\frac{\alpha_{20}}{2})
       }{
       1+K^2/2+(\alpha-\theta)^2\gamma_0^2}
       \!\!\!\!\!\!\!\!\!\!\!\!\!\!\!\!\!\!\!\!\!\!\!\!\!
       &&
  \\
  \phantom{f_{\Delta\psi_1}}
  \approx
  \frac{\delta_1\,\alpha_{10}(\alpha_{10}+\alpha_{20})(\theta-\alpha)}
       {2(1+K^2/2+(\alpha-\theta)^2\gamma_0^2)}
  &&
\end{eqnarray}
\begin{eqnarray}
  \nonumber
  &&
  f_{\Delta\Omega_1} {\ }={\ }   f_{\Omega_0}
  \; \frac{1+K^2/2}{1+K^2/2+\gamma^2(\alpha-\theta)^2}
  \;+
  \\ \nonumber && \quad
  \;+\; \frac{\delta_1\gamma_0^2(\alpha-\theta)^2}{c(1+K^2/2+\gamma^2(\alpha-\theta)^2)}
  \cdot
    \\ \nonumber && \quad
       (\cos\alpha_{10}-1-\sin\alpha_{10}\tan\frac{\alpha_{20}}{2})
      \; B_0B_1
   \\ \nonumber && \quad
   \approx \;
   \frac{\gamma_0^2(\theta-\alpha)\delta_1\alpha_{10}(\alpha_{10}+\alpha_{20})}
        {2c(1+K^2/2+(\alpha-\theta)^2\gamma_0^2)}
   \,B_0B_1
   \\  && \quad
   - \;
   \frac{(\alpha_{20}+\delta_2)(\alpha_{10}+\alpha-\theta)}
        {2c(1+K^2/2+(\alpha-\theta)^2\gamma_0^2)}
   \,B_1B_2
\end{eqnarray}
\begin{eqnarray}
\nonumber
  f_{\Delta x_0} &=&
  \frac{\nu_L}{c}{\,}
  \delta_1
  (\cos\alpha_{10}-1-\sin\alpha_{10}\tan\frac{\alpha_{20}}{2})
  \\ & \approx &
  \frac{\nu_L}{2c}{\,}
  \delta_1
  \alpha_{10}(\alpha_{10}+\alpha_{20})
\end{eqnarray}
\begin{eqnarray}
  &&
  \nonumber
  \!\!\!\!\!\!\!\!\!\!\!\!\!\!\!\!\!\!\!\!\!\!\!\!\!\!\!\!\!\!\!\!
  f_{\Delta \alpha_0} {\ }={\ }
  \frac{\nu_L}{c} \,
  \frac{\alpha_{10}(\theta-\alpha)\gamma_0^2\tan\frac{\alpha_{20}^2}{2}}{1+K^2/2}
  \,B_1B_2
  \\&&
  \!\!\!\!\!\!\!\!\!\!\!\!\!\!\!\!\!\!\!
  \approx\;
  \frac{\nu_L}{c} \,
  \frac{\alpha_{10}(\theta-\alpha)\gamma_0^2{\alpha_{20}^2}}{2(1+K^2/2)}
  \,B_1B_2
\end{eqnarray}

The phase shift of
Eq.~(\ref{Delta-psi-D-Delta-psi-D-ref-detuning-appendix})
provides the free-electron lasing without inversion if the
coefficients $f_{\Omega_0}$, $f_{\Delta\Omega_1}$, and
$f_{\Delta\psi_1}$ ensure the negative dispersion and the
sensitivity of the phase shift $\Delta \psi_D$ to the
wiggler-induced changes of the resonant detuning and phase of
Eq.~(\ref{Delta-psi-D}):

\begin{eqnarray}
 \label{condition-chi-Omega0-appendix}
 && f_{\Omega_0} \;=\; \chi \, L_W/c
\\
 \label{condition-chi-DeltaOmega1-appendix}
 && f_{\Delta\Omega_1} \;=\; (\chi+\xi) \, L_W/c
\\ \label{Delta-psi-D-f-requirements-appendix}
 && f_{\Delta\psi_1} = \vartheta
 \,.
\end{eqnarray}

These matching conditions determine the drift region geometry: the
lengths $L_1$ and $L_2$ and the angle $\delta_1$, which are
specified by
Eqs.~(\ref{negative-detuning-geometry-condition})-(\ref{drift-region-sensitivity-delta1-condition}).
The condition for compensation of the drift-region induced
divergence determines the inhomogeneity length $b$ of the second
magnet TM2~(\ref{b-divergence-compensate-estimate}).


\begin{thebibliography}{99}

\bibitem{long-wavelength-FELs}
H.P. Freund and V.L. Granatstein,
Nuclear Instr. \& Meth. Phys. Res. A $\bf 429$, 33 (1999);

\bibitem{short-wavelength-FELs}
B. Sonntag,
Nuclear Instr. \& Meth. Phys. Res. A $\bf 467-468$, 8 (2001);



\bibitem{FELs-scientific-applications} M.E. Couprie, J.M. Ortega,
``Free-electron lasers sources for scientific applications'',
Analusis {\bf 28}, 725-736 (2000);

\bibitem{FEL-quantum-wells-spectrosopy}
T. Asano, S. Noda, A. Sasaki, T. Suzuki, T. Mitsuyu, K. Nishi, H.
Ohyama, T. Tomimasu, ``Ultrafast interband-light modulation by
intersubband light (free-electron laser) in quantum wells'', Nucl.
Inst. \& Meth. Phys. Res. B {\bf 144}, 123-129 (1998);


\bibitem{nonlinear-spectroscopy-of-solid-films}
G.N. Zhizhin, E.V. Alieva , L. Kuzik, V.A. Yakovlev, D.M.
Shkrabo, A.F.G. van der Meer, M.J. van der Wiel, ``Free-electron
laser for infrared SEW characterization surfaces of conducting
and dielectric solids and nm films on them'', Appl. Phys. A
--- Mater. Sc. \& Proc {\bf 67}, 667-673 (1998);


\bibitem{near-field-surface-microscopy}
A. Cricenti, R. Generosi, P. Perfetti, G. Margaritondo, J.
Almeida, J.M. Gilligan, N.H. Tolk, C. Coluzza, M. Spajer, D.
Courjon, I.D. Aggarwal, ``Interface applications of scanning
near-field optical microscopy with a free electron laser'',
Physics Status Solidi A -- Applied Research {\bf 175}, 317-329
(1999);

\bibitem{Molecular-spectroscopy-FELs}
M. Putter, G. von Helden, G. Meijer, Chem. Phys. Lett., ``Mass
selective infrared spectroscopy using a free electron laser''
 {\bf 258}, 118-122 (1996);

\bibitem{Laser-surgery}
J.T. Payne, G.M. Peavy, L. Reinisch, et al., ``Cortical bone
healing following laser osteotomy using 6.1 mu m wavelength'',
Laser Surg. Med. {\bf 29}, 38-43 (2001);

\bibitem{Laser-surgery-eye}
V.C. Coffey, ``Free-electron lasers - Eye surgery application
demonstrated with FEL'', Laser Focus World {\bf 37}, 59-60 (2001);

\bibitem{FEL-laser-technology}
P.G. O'Shea, H.P. Freund, ``Laser technology --- Free-electron
lasers: Status and Applications'', Science {\bf 292 (5523)},
1853-1858 (2001);

\bibitem{FEL-X-ray-matter-interaction}
S. Doniach, ``Fourth-generation X-ray sources: some possible
applications to biology'', J. Synchrotron Radiat. {\bf 7} 116-120,
(2000);

\bibitem{FEL-emittance}
S. Reiche,
Nuclear Instr. \& Meth. Phys. Res. A {\bf 445}, 90 (2000);

\bibitem{LWI} M.O. Scully, S.Y. Zhu, and A. Gavrielidis,
``Degenerate quantum-beat laser: Lasing without inversion and
inversion without lasing'', Phys. Rev. Lett. {\bf 62}, 2813
(1989);

\bibitem{1} G. Kurizki, M.O. Scully, and C. Keitel, Phys. Rev. Lett. {\bf 70}, 1433
(1993);

\bibitem{2} B.Sherman and G. Kurizki, Phys. Rev. Lett. {\bf 75}, 4602
(1995);

\bibitem{3} D.E. Nikonov, B. Scherman, G. Kurizki, M.O. Scully, Opt. Commun. {\bf 123}, 363
(1996);

\bibitem{4} D.E. Nikonov, M.O. Scully, and G. Kurizki, Phys. Rev. E {\bf 54}, 6780
(1996);

\bibitem{5} D.E. Nikonov, Yu.V. Rostovtsev, and G. Sussmann, Phys. Rev. E {\bf 57}, 3444
(1998);

\bibitem{6} A.I. Artemiev, M.V. Fedorov, Yu.V. Rostovtsev, G.
Kurizki, and M.O. Scully, Phys. Rev. Lett. {\bf 85}, 4510 (2000);

\bibitem{7} Y.V. Rostovtsev, G. Kurizki, M.O. Scully,
``Broadband optical gain via interference in the free electron
laser: Principles and proposed realizations'', Phys. Rev. E
 {\bf 64}, 026501 (2001);

\bibitem{Brau}
Free-Electron Lasers, C. A. Brau, Academic Press, Boston (1990);

\bibitem{Scully-1976}
F.A. Hopf, P. Meystre, M.O. Scully, W.H. Louisell, Phys. Rev.
Lett. {\bf 18}, 413 (1976);

\bibitem{Datolli-PRE-V61-p7052} G. Datolli, L. Mezi, L. Bucci,
Phys. Rev. E {\bf 61}, 7052 (2000);

\bibitem{Madey}
J. M. J. Madey, Nuovo Cimento B {\bf 50}, 64 (1978);


\bibitem{ring-megative-dispersion}
M. Giovannozzi, B. Autin, M. Chanel, M. Martini, Ph. Royer,
``Application of wigglers to quasi-isochronous transport
systems'', Proceedings of EPAC 2000, \newline \noindent
http:$\!$/$\!\!$/accelconf.web.cern.ch/AccelConf
\newline
/e00/PAPERS/TUP6B09.pdf{$\;$};
\newline \noindent
S. Y. Lee, K. Y. Ng, and D. Trbojevic, ``Minimizing dispersion in
flexible-momentum-compaction lattices'', Phys. Rev. E {\bf 48},
p. 3040 (1993).


\end{thebibliography}
\end{document}